\documentclass[lettersize,journal]{IEEEtran}

\usepackage{amsmath,amsfonts}
\usepackage{algorithmic}
\usepackage{algorithm}
\usepackage{array}
\usepackage[caption=false,font=normalsize,labelfont=sf,textfont=sf]{subfig}
\usepackage{textcomp}
\usepackage{stfloats}
\usepackage{url}
\usepackage{verbatim}
\usepackage{graphicx}
\usepackage{cite}
\usepackage{xcolor}
\usepackage{multirow}

\usepackage{amssymb}

\begin{document}
\newcommand{\DU}[0]{Dialogue Understandability}
\newcommand{\IF}[0]{Influencing Factor}

\title{Dialogue Understandability: \\ Why are we streaming movies with subtitles?}

\author{Helard Becerra, Alessandro Ragano, Diptasree Debnath, Asad Ullah, Crisron Rudolf Lucas, Martin Walsh, Andrew Hines,~\IEEEmembership{Senior Member,~IEEE,}
\thanks{This publication has emanated from research conducted with the financial support of Xperi Inc. and Science Foundation Ireland (SFI) under Grant Number 12/RC/2289\_P2. For the purpose of Open Access, the author has applied a CC BY public copyright licence to any Author Accepted Manuscript version arising from this submission.}
\thanks{H. Becerra, A. Ragano, D. Debnath, A. Ullah, C. R. Lucas, and A. Hines are with the School of Computer Science, University College Dublin,  Dublin, Ireland e-mail: helard.becerra@ucd.ie.}
\thanks{{M. Walsh is with Xperi Inc.,  San Jose, California e-mail: Martin.Walsh@xperi.com.}}
}



\maketitle

\begin{abstract}
Watching movies and TV shows with subtitles enabled is not simply down to audibility or speech intelligibility. A variety of evolving factors related to technological advances, cinema production and social behaviour challenge our perception and understanding. This study seeks to formalise and give context to these influential factors under a wider and novel term referred to as Dialogue Understandability. We propose a working definition for Dialogue Understandability being a listener’s capacity to follow the story without undue cognitive effort or concentration being required that impacts their Quality of Experience (QoE). The paper identifies, describes and categorises the factors that influence Dialogue Understandability mapping them over the QoE framework, a media streaming lifecycle, and the stakeholders involved. We then explore available measurement tools in the literature and link them to the factors they could potentially be used for. The maturity and suitability of these tools is evaluated over a set of pilot experiments. Finally, we reflect on the gaps that still need to be filled, what we can measure and what not, future subjective experiments, and new research trends that could help us to fully characterise Dialogue Understandability.
\end{abstract}


\section{Introduction}

Do you always have the subtitles on when watching Netflix? If so you are amongst the 85\% of Netflix users who turn on closed captions when streaming movies or TV series~\cite{kim2023comparing}. Closed captions or subtitles have been in use since the 1970s and are an invaluable accessibility tool for deaf or hard-of-hearing individuals~\cite{ellcessor2012captions}. They are also commonly used for foreign content or even in noisy environments, such as rolling news channels in airports.

Ideal Insight carried out a survey and found that 85\% of Netflix (54\% Amazon Prime and 37\% Disney+) customers opt to watch with captions~\cite{idealinsight}. {Another survey from YouGov reported that younger people seemed to enable subtitles more often than older people (63\% of adults under 30 years old)~\cite{yougov2023subtitles}. Such high percentages cannot be explained based on hearing impairments, age, language proficiency or noisy environments, so why are so many switching on subtitles?} The answer lies in the fact that streaming has changed not just what we watch, but how it is made, distributed and consumed. Can we identify, explain and potentially quantify the effects of these factors?

In this study, we explore the wide range of factors that can influence whether we follow the story or get frustrated and potentially give up on watching a movie. {We focus the analysis on factors related to 1) the production of movies (e.g., sound recording, actor speaking style, etc), 2) the distribution of movies (e.g.,  sound coding, streaming quality), and 3) the listener environment (e.g., room acoustics, tv integrated speaker, headphones).} We introduce the concept of \DU{} as a term to encompass more than just intelligibility or basic word understanding. We will use the term movie as a generic term for content, but the concept could equally be applied to any cinematic, TV or streaming content. {We explore \DU{} from a universal perspective, acknowledging that many of the issues and factors are further exacerbated for individuals with hearing impairments.}

In this paper, we propose a working definition for \DU{} in the context of streamed content. To support this, we review, {identify} and categorise the factors that impact \DU{} (Section~\ref{sec:background}). We show the links between the defined \DU{} factors and the established Quality of Experience (QoE) framework, the lifecycle stages of media streaming, and the involved stakeholders (Section~\ref{sec:DU_Factors}). We explore potential methods for automatic characterisation of \DU{} factors and present a set of pilot studies to illustrate the feasibility of applying these methods over movie-like content (Section~\ref{sec:tools}). Finally, we reflect on \DU{} and its utility to characterise the influence of sound and subtitles on the QoE of our movie consumption (Section~\ref{sec:discussion}).

\section{\DU{}}\label{sec:du}

\subsection{Working Definition}
Several speech and dialogue aspects (e.g., quality, intelligibility, content, etc.) have been defined and studied over the years, resulting in well-defined assessment procedures. However, \DU{} for movies is a broader concept. For instance, the traditional concept of intelligibility is usually measured as the listener's capacity to hear words as a percentage of the overall dialogue~\cite{schepker2016perceived}. Yet, over a streamed movie session, high levels of intelligibility might not guarantee a viewer's complete understanding of the story plot. Using a traditional speech quality approach to describe and measure \DU{} would also fall short, as the levels of degradation captured by speech metrics might not be entirely relevant to measuring the engagement of users and their perception of quality~\cite{dobrian2011understanding}.

In this study, the term \DU{} is used in the context of cinematic, TV and streamed content and is defined as follows:

\begin{center}
  \textit{"The listener's capacity to follow a story plot without undue cognitive effort or concentration being required which impacts the Quality of Experience."}
\end{center}

\subsection{Root causes}

\DU{} is constantly exposed to different factors that affect it at different levels. In~\cite{gorman2021adaptive}, these factors are referred to as situational factors, such as mumbling actors, sound effects/content background noise, loud consumption environments (e.g., metro, bus, crowded street), lexical complexity, languages/accents, and consumption habits. Some of these factors were also reported as anecdotal examples in video articles that highlighted the use of subtitles as a means to improve \DU{}~\cite{vox2023}.

\begin{figure}
    \centering
    \includegraphics[width=0.44\textwidth]{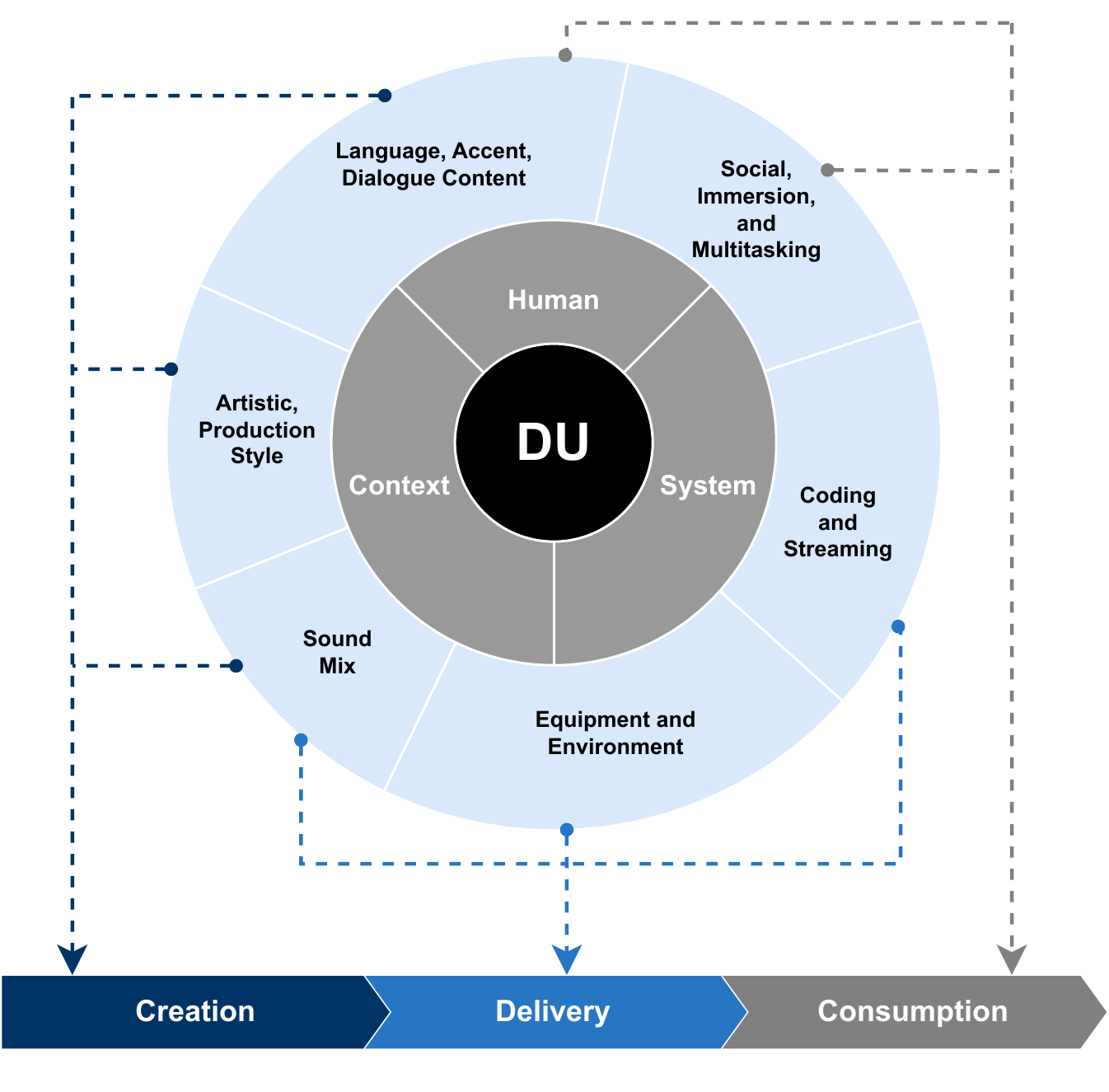}
    \caption{The \DU{} \IF{}s mapped to the QoE \IF{}s .}
    \label{fig:QoEMapping}
\end{figure}

Although a range of studies analyses some of these factors individually from different angles\cite{shirley2021intelligibility, tapaswi2016movieqa, zhu2015understanding}, to our knowledge, there is no formal study listing and categorising factors affecting movie consumption at their different stages. To fill this gap, we explore the root causes, identify the main influential factors and organise them into six sub-categories: 1)~artistic and production style, 2)~language, accents and dialogue content, 3)~sound mixing techniques, 4)~coding and streaming technology, 5)~equipment and consumption environment, and 6)~social, immersion and multitasking. Figure~\ref{fig:QoEMapping} presents these categories mapping them to the influential factors defined in the QoE framework (human, system, and context)~\cite{QualinetWP:2013}. This mapping process was inspired by previous studies that used the QoE framework to organise and facilitate the analysis for tasks like the quality assessment of sound archives and the listening effort analysis~\cite{ragano2022automatic, 10.3389/fpsyg.2021.767840}. \DU{} factors are also traced back to three main stages in a common movie consumption pipeline (creation, delivery, and consumption). Figure~\ref{fig:QoEMapping} illustrates the spectrum of factors affecting streaming media consumption, linking them to the lifecycle stages of movie consumption. {Using a lifecycle pipeline for analysis was done previously in a study presented by Ragano et al.~\cite{ragano2022automatic}. to analyse and assess the perceived quality of restored sound archives throughout different lifecycle stages (digitization, restoration, and consumption).}

\section{Related Concepts}\label{sec:background}

Here, we review concepts that are related to aspects of \DU{} but do not individually capture it. We briefly define each concept, their relevance in the context of \DU{}, and how they are measured (summary in Table~\ref{tab:basic-concepts}).

\begin{table*}[tb]
  \centering
  \caption{Summary of related concepts affecting \DU{} (DU).}
  \resizebox{2\columnwidth}{!}{%
    \begin{tabular}{lllp{30.855em}}
    \hline
    \textbf{Concept} & \textbf{Media} & \textbf{Aim} & \multicolumn{1}{l}{\textbf{Limitations}} \\
    \hline
    QoE   & Audio, speech, video, image, text & Provides conceptual framework & \multicolumn{1}{l}{DU not systematically studied within QoE framework} \\
    Speech quality & Speech & Assess listening experience & Designed for speech communication.\newline{}Not capturing artistic choices in entertainment content \\
    Speech intelligibility & Speech & Measure word recognition rates & \multicolumn{1}{l}{Intelligibility levels don't always correspond to DU levels} \\
    Loudness & Audio, speech & Measure sound pressure levels (intensity) & Determines the dynamic range and affects the dialogue loudness.\newline{}Not sufficient to describe DU \\
    Speech rate & Speech & Measure words per second & Affects perception of clarity, comprehension and fluency of speech.\newline{}Not sufficient to describe DU \\
    Speaker identification & Speech & Identification based on speech characteristics & Identify actors that are more challenging to understand.\newline{}Not sufficient to describe DU \\
    Immersion & Audio, speech, video, image, text & Measure cognitive attention & \multicolumn{1}{l}{Not sufficient to describe DU} \\
    Multimodal interactions & Audio, speech, video, image, text & Assess quality perception of multimodal media & Explores complex interactions in dialogue and visual cues.\newline{}Not sufficient to describe DU \\
    \hline
    \end{tabular}%
    }
  \label{tab:basic-concepts}%
\end{table*}%

\subsection{Quality of Experience (QoE)}
\label{sec:qoe}
QoE is defined as \textit{the degree of delight or annoyance of the user of an application or service. It results from the fulfilment of his or her expectations with respect to the utility and/or enjoyment of the application or service in the light of the user’s personality and current state}~\cite{QualinetWP:2013}. QoE uses a conceptual framework and categorises the influencing factors affecting how users perceive the quality of a system or service into three classes: human, system, and context (see Table \ref{tab:qoe}). This provides researchers with a consistent yet flexible framework they can use to guide the analysis and assessment of services and applications following a user-centric approach. Although several studies have targeted applications like audio-visual signal transmission, speech quality and intelligibility, cognitive and listening effort, etc~\cite{akhtar2019multimedia}, which add to the \DU{} concept, QoE analysis was commonly restricted to the delivery and consumption stages~\cite{qadir2015novel}, omitting the creation stage, where critical choices that can affect \DU{} are made. QoE measurement adopts two main approaches: subjective assessments that rely on human opinions and objective assessments that use computational algorithms to quantify the perceived quality.

\begin{table}[htbp]
  \centering
  \caption{Illustrative classification of Quality of Experience \IF{}s in the context of telecommunications from \cite{Reiter2014}.}
  \resizebox{1\columnwidth}{!}{%
    \begin{tabular}{rl}
    \hline
    \multicolumn{1}{l}{\textbf{Infuential Factors}} & \textbf{Examples} \\
    \hline
    \multicolumn{1}{l}{Human} & Low-level processing (visual and auditory acuity, gender, age, mood) \\
          & \multicolumn{1}{p{31.07em}}{Higher-level processing (cognitive processes, socio-cultural and \newline{}economic background, expectations, needs and goals, other personality traits)} \\
    \multicolumn{1}{l}{System} & Content \\
          & Media (encoding, resolution, sample rate) \\
          & Network (bandwidth, delay, jitter) \\
          & Device (screen resolution, display size) \\
    \multicolumn{1}{l}{Context} & Physical (location and space) \\
          & Temporal (time of day, frequency of use) \\
          & Social (inter-personal relations during experience) \\
          & Economic \\
          & Task (multitasking, interruptions, task type) \\
          & Technical and information (relationship between systems) \\
    \hline
    \end{tabular}%
    }
  \label{tab:qoe}%
\end{table}%

\subsection{Speech Quality}\label{subsec:bg_quality}
{Speech sounds can be distinguished on attributes like pitch, duration, loudness, timber, content and spaciousness. The sum of these components describes the perceived composition of the speech sound and its perceived quality~\cite{moller2011speech}.} On its own, speech quality is a broad concept related to the overall listening experience for a speech signal without focusing on the information that speech conveys~\cite{raake2007speech}. For instance, if speech is sampled at 8 kHz, it can still be understood, but its quality might be perceived as low. {Another major concern for speech quality is noise (or background noise), as its presence at the send side could lead to masking effects lowering speech intelligibility~\cite{raake2007speech}.} Speech quality assessment evolved for technology assessment, that is, evaluating speech codecs and telephone speech. This resulted in several methods for automatic assessment~\cite{moller2011speech}. However, content or aspects like artistic choices or actor voices are not part of the scope of these methods. For example, the voice of Darth Vader in Star Wars may be incorrectly labelled as poor quality when, in fact, it represents an intentional artistic decision. Thus, it could be argued that despite the speech quality capacity to describe and measure some aspects affecting \DU{}, the current objectives of speech quality assessment clash with the artistic intentions underlying the creation of movies.

\subsection{Speech Intelligibility}\label{subsec:bg_intelli}
{Speech intelligibility measures how the content of a speech utterance is correctly identified, that is, the number of words or phonemes correctly recalled in a speech sample~\cite{moller2011speech}.} Ratings can be obtained through listening tests such as the speech reception threshold (SRT)~\cite{plomp2001intelligent} or the word error rate (WER)~\cite{levenshtein1966binary}. These measures, combined with voice activity detection (VAD) algorithms,  are important in speech communications, especially for individuals with speech and hearing impairments and communication systems where speech intelligibility must be preserved. Speech intelligibility, while a contributing factor to \DU{}, is not sufficient on its own to fully quantify it. A high or low intelligibility rating does not necessarily correspond to high or low \DU{}. In fact, it can be argued that even if certain words are unclear, the overall meaning of a movie dialogue can still be grasped. This situation is commonly observed in interactions between native and non-native speakers, where native listeners rely on contextual cues to comprehend conversations and anticipate sentences~\cite{lev2015comprehending}. {Conversely, high levels of intelligibility might not always guarantee \DU{}, for instance, it has been pointed out that intelligibility does not account for irony or sarcasm which might be critical to understanding context~\cite{puhacheuskaya2022being}.}

\subsection{Loudness} \label{subsec:bg_loudness}
Loudness refers to the subjective perception of the sound pressure level. It is expressed in decibels (dB) and can be measured using specialised equipment such as sound level meters or acoustic analysis software. The role of loudness is a highly influencing factor in movies. {It will affect the speech-to-background ratio (SBR) which measures the balance between dialogue and non-speech elements in a sound mix audio track; and it will also determine the dynamic range which is defined as the intensity range between the softest and the loudest sound within a film scene.} According to Netflix, content that avoids excessive dynamic range provides a better experience for customers~\cite{netflixmix}. At its encoding stage, Netflix performs loudness normalisation for the movies’ audio master tracks where speech activity is detected~\cite{medium2013netflix}. Dialogue loudness can be a determinant factor for \DU{}, as mumbling actors have been identified as one of the main factors that contribute to the reliance on subtitles~\cite{theguardian}. Computational algorithms to measure speech loudness have been developed over the years~\cite{mittag2021nisqa}, while it can help describe key aspects in \DU{}, it alone is not sufficient to fully address it.

\subsection{Speech Rate} \label{subsec:bg_speakerrate}
Speech rate refers to the speed or pace at which an individual speaks, typically measured in terms of words per minute (WPM), syllables per second (SPS) or phonemes per second (PPS)~\cite{yuan2006towards}. Speech rate can be measured by analysing the duration of an audio clip or a specific speech segment and counting the number of words, syllables or phonemes within that timeframe~\cite{de_jong_praat_2009}. Alternatively, automatic speech recognition (ASR) systems can be used to transcribe speech and estimate speech rate based on the recognised words. Within movies, speech rate can vary depending on different aspects like the character's role, an actor's intrinsic characteristic or the scene context. High speech rate could increase cognitive effort and the ability of viewers to keep up with the subtitles, which, in turn, might affect \DU{}.

\begin{figure*}[tb]
    \centering    \includegraphics[width=1\textwidth]{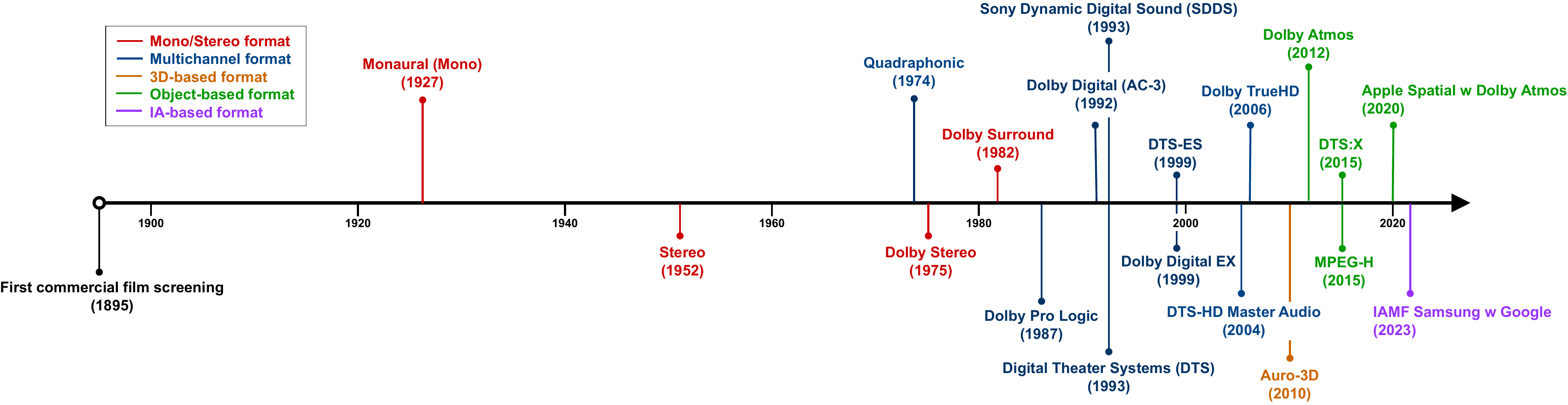}
    \caption{Evolution of sound mixes from 1920 to 2020.}
    \label{fig:evolution_mix}
\end{figure*}

\subsection{Speaker Identification}\label{subsec:bg_speakerid}

Speaker identification aims to determine the identity of an individual based on their voice and speech characteristics~\cite{fan2013acoustic}. The application of ML-based techniques have improved the accuracy of speaker identification models~\cite{jahangir2021speaker}. {From a \DU{} standpoint, speaker characteristics are important as some actors may be more challenging to comprehend than others. A recent survey conducted in the United States revealed that the top three actors who were reported as difficult to understand were Tom Hardy, Sofia Vergara, and Arnold Schwarzenegger~\cite{survey_america}.} Difficulties and increased cognitive load associated with certain actors can also arise from language variations, for instance, American-English viewers listening to British-English actors or foreign accents. Being able to identify speakers within movies might help spot potential issues related to \DU{} and could guide the implementation of additional assistance tools to ensure \DU{}. {For instance, personalised dialogue enhancement algorithms or selectively enabled subtitles, where users could select which character(s) or language(s) to target.}

\subsection{Immersion}\label{subsec:bg_immersion}
Immersion can be defined as \textit{the cognitive response of the user to the characteristics of a system or media content}~\cite{nilsson2016immersion}. Several studies have explored how to measure the levels of immersion of a subject for a given stimulus. The Qualinet’s White Paper on Definitions of Immersive Media Experience (IMEx)~\cite{perkis2020qualinet} narrowed the methods to measure immersive media experience to three main categories: subjective, behavioural, and psycho-physiological assessments. For \DU{}, the levels of immersion experienced during a movie session can determine whether a user is able to follow or not the content’s plot, which will determine the user engagement (whether the user gets frustrated and stops watching that movie). That immersion is partially shaped by technological advances and evolving consumption habits (e.g., watching a movie on a smart TV while sending an instant message over a smartphone). Such a level of immersion can also be affected by other elements like the screen size, the environment, or narrative factors. {From a sound mix perspective, sound formats for home and cinema have evolved, expanding the number of channels available and adding new components to deliver a more immersive movie-watching experience for consumers~\cite{yuan2014beyond}. Figure~\ref{fig:evolution_mix} depicts the evolution of different sound mix techniques over time, going from Mono/Stereo, multichannel, 3D and object-based formats. IA-based formats are now being explored to fit consumers and the way they perceive sound~\cite{nam2021ai}.} Being able to measure such levels of immersion (subjective experiments) is critical to describe \DU{}.

\subsection{Multimodal Interactions}\label{subsec:bg_multimodal}


Research on multimodal interaction explores how different types of media affect each other and, in turn, the overall perceived quality~\cite{yuan2014beyond, dupont2000audio}. Audio and visual components are the most common modalities for media content with several studies exploring their mutual influence and their contribution to the overall quality perception~\cite{akhtar2017audio, becerra2021perceptual}. However, multimodal research is not restricted to just these two components. Subtitles (closed captions) are traditionally included in audio-visual content to assist viewers with hearing loss while they consume specific media content~\cite{lewis1999television}. For \DU{}, the interaction between these components gets more complex with the inclusion of techniques like shot-cuts between scenes, speakers off-camera, or visual cues to prompt speech intelligibility. In a video article, the portal Vox provides compelling examples of why we have an increasing reliance on subtitles to keep up with complex movie plots, unintelligible dialogue sequences, and {sound mixing limitations}~\cite{vox2023}.

\section{\IF{}s}\label{sec:DU_Factors}

The six categories of factors affecting \DU{} presented in this study are mapped to the human, system, and context QoE factors in Figure~\ref{fig:QoEMapping}. Classifying the \DU{} factors using the QoE framework will assist in identifying and characterising methodologies and tools to quantify \DU{} for movie content. We also consider how these factors interact temporally across the different stages of a movie lifecycle. We break the lifecycle into three main stages:

\begin{enumerate}
    \item \textit{Creation:} Covers the production stages related to film development, pre-production, film shooting, and post-production \cite{swartz2005understanding};
    \item \textit{Delivery:} Considers the processes by which media is encoded, transmitted and adapted to be reproduced over a particular device \cite{moller2014quality};
    \item \textit{Consumption:} Covers the actual reproduction of the media by the consumer \cite{zhu2015understanding}.
\end{enumerate}



\begin{figure*}[tb]
    \centering    \includegraphics[width=.9\textwidth]{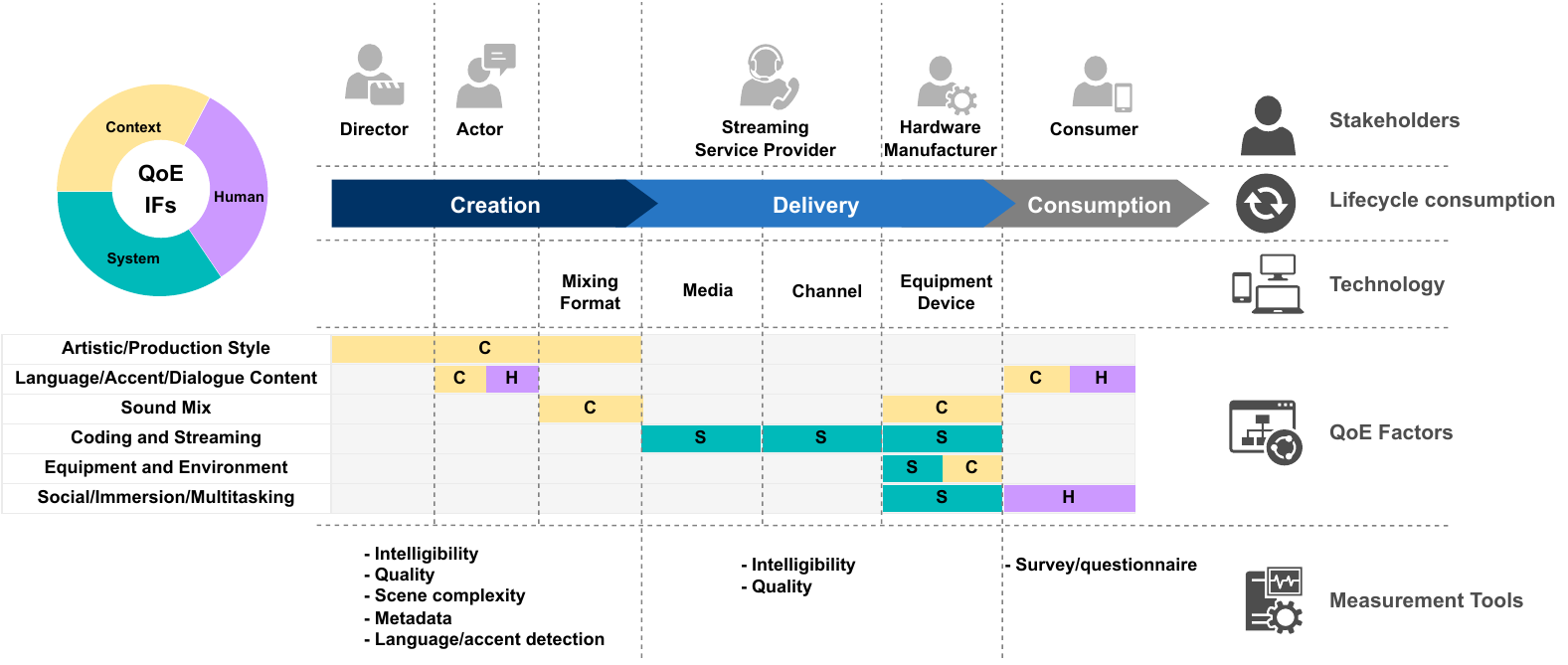}
    \caption{The \DU{} lifecycle and key stakeholders. The QoE \IF{}s (top left) are mapped to stages of the lifecycle (creation, delivery, consumption) for each of the \DU{} \IF{}s. Key stakeholders and systems at each stage are also illustrated.}
    \label{fig:PipelineMapping}
\end{figure*}

Figure~\ref{fig:PipelineMapping} illustrates how \DU{} is influenced across the stages of a movie pipeline and can be broadly grouped into six factors. It also captures the stakeholders, technologies and measurement tools, and clusters the factors by QoE classes: human, context and system. This helps us to identify complex interactions during the movie lifecycle; for instance, we can see that the language, accent and dialogue content can be considered as both human and context factors. We can also see that \DU{} factors are not restricted to one unique stage in this lifecycle. For instance, sound mixing at the creation stage can affect \DU{}, where a sound engineer combines the different sound elements of the movie, but it can also occur at delivery stage, where the sound mix is adapted (downmixed) to a sound mix format of a particular consumption device. {Below, each of the \DU{} influencing factors are briefly described using illustrative examples.}

\subsection{Artistic/Production Style}\label{subsec:art_prod_style}
Audio-visual production has advanced in realism since broadcast TV of the 20th Century. Long-running TV shows like Doctor Who or Star Trek illustrate the evolution from an almost theatrical stage play with dialogue spoken to camera to immersive dialogue on the run \cite{makingwaves2019}. The artistic/production style factors were mapped to a context QoE factor at the creation stage (Figure~\ref{fig:PipelineMapping}). At this stage, content creators (producers, writers, directors) will come up with the idea for the intended content and bring their artistic perspective to the project (film, TV show, documentary, etc.). During this stage, the content idea, script draft, the desired film locations and the artistic and creative guidelines are defined. Despite the effect that they could have on the \DU{}, artistic/production guidelines are crucial to creators as they consider them as key to reflect the intended ‘meaning’ they try to imprint on the content and the desired experience they want to deliver to the target audience~\cite{QualinetWP:2013}. Artistic and production style features can be sourced out through metadata available (e.g., genre, director, cast, etc). Other speech and visual informative features could be obtained using automatic methods~\cite{mittag2021nisqa, union2008itu}.

\subsection{Language, Accents and Dialogue Content} \label{subsec:lang_acc_cont}
War of the Worlds~\cite{waroftheworlds2019} is an example of a multilingual series co-produced by Canal+, Fox and Star where, unless you are fluent in several languages, you will need some subtitles. Audio-visual media regulations~\cite{europaurl}, streaming companies' investment strategies~\cite{limov2020click}, and the desire for natural dialogues have resulted in national productions with multiple languages and/or strong local accents that people are not familiar with. The language, accent and dialogue content factors were mapped to both context and human QoE factors, and they were traced to the creation and consumption stages at the streaming lifecycle (Figure~\ref{fig:PipelineMapping}). At the creation stage, the context is guided by local content regulations, investment strategies~\cite{europaurl,limov2020click} and the need for more natural and realistic productions. The human factor is set by the actors’ speech characteristics and their capacity for delivering natural dialogues (e.g., accent and dialect mastery). At the consumption stage, the context is determined by the area/country where the content is being consumed (e.g., English-language content over non-anglophone regions)~\cite{sanchez2021netflix}. In turn, the human factor is determined by the consumers’ traits. Being a native speaker (L1) or having a second language (L2) would put a consumer in a better position while engaging with certain content. Whether someone sticks with a movie to the end or continues to watch a series is influenced by more than the plot. Understanding the effect of language, accent, and dialogue content from a \DU{} perspective will help identify regions where subtitles or dubbing would make a difference in the level of consumption and acceptance. {It can also guide the implementation of personalised tools for dialogue enhancement or smart subtitling for specific languages or accents.} Language and accent features can be outsourced through the metadata available. More recently, progress in ML-based techniques have allowed the use of automatic tools for language and accent recognition~\cite{Zuluaga23}.

\subsection{Sound Mix} \label{subsec:sound_mix}
Film director Christopher Nolan (Tenet, Interstellar) is well known for his special effects but also for his sound mixing. Some dialogue is meant to be unintelligible and difficult to understand, but it is hard for the audience to know whether this is intentional or not~\cite{guardiancnolan}. {The balance between dialogue and non-speech elements in a sound mix, the Speech-to-Background Ratio (SBR), has a great impact on people’s ability to discern dialogue, especially for hearing impaired consumers. Experiments highlighted the variance of ratio preferences among subjects and content type (e.g., news, sports, drama)~\cite{torcoli2019background}.} The sound mixing factor was mapped to a context QoE factor and traced to the creation and delivery stages of the streaming lifecycle (Figure~\ref{fig:PipelineMapping}). At the creation stage, content creators (directors and producers) choose certain mixing strategies and format standards that they consider important for transmitting certain experiences. During the delivery stage, hardware manufacturers must provide compatible devices to adapt the mixing formats available. The evolution of microphones and sound recording equipment resulted in several directors guiding their sound mixing decisions to target primarily the widest surround sound formats available (e.g., Dolby Atmos allowing up to 128 audio tracks). From a \DU{} standpoint, these choices come at the cost of decreased intelligibility for users that consume content using mobile devices or standard home sound mixing settings (e.g., 5.1, stereo and mono mixing formats). {At the creation stage, mixing engineers often have access to the script or have listened to the dialogue several times. For them, this could lead to content familiarity and reduce their sensitivity to poor \DU{}. For dubbed material, streaming service providers often require material for several languages. Due to time and budget constraints, there is often minimal or no quality assurance on how the new language melds with the original stem.} Sound mixing characteristics are commonly available through metadata and the technical specs from the content. Collecting this information can help guide the implementation of dedicated audio processing assistance tools for consumption devices to improve the consumers’ experience (e.g., dialogue enhancement fitted for specific content and devices)~\cite{torcoli2019background}.

\subsection{Coding and Streaming}\label{subsec:codecs_streaming}
With the rise of the over-the-top (OTT) model, which aims to deliver movies and TV content via the Internet, streaming companies are now compelled to provide their customers with services and equipment that guarantee high levels of perceived quality in order to keep themselves as relevant alternatives in the OTT market. The coding and streaming factor was mapped to the QoE system factor, and it was traced to the delivery stage in the streaming lifecycle (Figure~\ref{fig:PipelineMapping}). At the delivery stage, system factors involve the media characteristics (e.g., video frame rate, sound mixing setup), the transmission channel characteristics (e.g., the bandwidth and network characteristics), and the equipment and device characteristics (e.g., high-resolution smartphone)~\cite{QualinetWP:2013,akhtar2017audio}. Streaming service providers are responsible for storing the media content in data centres equipped with high-speed internet connection for efficient distribution. Meanwhile, hardware manufacturers must provide devices that adhere to coding, network and transmission standards for audio playback. Tools to compute coding and streaming features have received much attention from the research community with several objective tools being developed and standardised throughout the years~\cite{citation-0}. They have served as a base for the implementation of adaptive streaming technology to automatically tune content quality based on the users’ network conditions, thus ensuring a better experience without disruption in reproduction~\cite{chen2015smart,zhou2016mdash}.

\subsection{Equipment and Environment}\label{subsec:equip_env}
Advances in the consumer electronic sector have shaped the way we consume and experience movies outside cinema theatres. For movie consumption, users now have at their disposal a large variety of devices (e.g., smartphones, personal computers, smart TVs, gaming consoles, home theatre systems, etc.), each one supporting different sound mixing formats (e.g., mono, 2.0 stereo, 5.1 and 7.1 surround), making the sound mixing adaptation process harder. {Modern TV manufacturers aim for thinner designs with small speakers and paper-thin drivers. To compensate for these drivers, a combination of psychoacoustics and digital signal processing methods are commonly used to make the sound perceived as louder, affecting the dialogue intelligibility.} The equipment and environment factors were mapped to system and context QoE factors, and they were traced to the delivery stage in the streaming lifecycle (Figure~\ref{fig:PipelineMapping}). At the delivery stage, the system factor is guided by hardware manufacturers and their engagement to meet the wide range of requirements for sound downmixing and format adaptation. Meanwhile, the context is guided by the type, complexity, and usability of new consumption devices (e.g., mobility provided by smartphones, tablets and other portable devices)~\cite{falkowski2020consumption}. Sound downmixing and format adaptation for consumption devices pose several challenges to guarantee good levels of \DU{}. Sound elements, potentially key for a dialogue sequence, might get lost or overlapped due to the downmixing process. Acquiring expensive equipment (e.g., high-quality soundbars) for a better experience might not be a feasible solution if companies are interested in reaching large audiences. Device mobility of new consumption devices comes at the expense of environmental noise and distractions, which, depending on the service and the content being consumed (e.g., a phone call in the street, music while running, movies at the metro), could affect \DU{}~\cite{falkowski2020current}. {How and where the content is consumed affects \DU{}, this could explain why people are much less likely to enable subtitles when listening with headphones.} Equipment and environmental information might not be available at the consumer end. Questionnaires over subjective experiments might be the best alternative to capture this information and analyse its effect over the perceived \DU{}.

\subsection{Social/Immersion/Multitasking}\label{subsec:social_immersion}
Social behaviour has evolved throughout the years, and so has the way we watch movies. Studies have reflected on multitasking and its effect while consuming streamed content (e.g., watching a movie while chatting on social media apps)~\cite{shokrpour2017people}. The social, immersion and multitasking factors were mapped to the system and human QoE factors, and they were traced to the delivery and consumption stages (Figure~\ref{fig:PipelineMapping}). At a delivery stage, hardware manufacturers are compelled to design devices that fulfil the needs of more demanding users and the so-called 'digital-native' generation~\cite{aagaard2015media}. At the consumption stage, the human factor is guided by users' technological demands (e.g., constant connectivity, on-demand services), consumption habits (e.g., watching content at home while doing chores) and social traits (e.g., watching a movie at night with low volume). {Social and cultural traits have also influenced the design of our living spaces, where media is commonly consumed. Modern homes have shifted from cluttered interiors to minimalist designs, often with bare walls, large windows, and hard floors. This acoustic transformation has introduced room modes, early reflections, and lingering echoes, hindering dialogue comprehension in living areas. Additionally, open-plan layouts expose viewers to household noises from appliances, often masking dialogue altogether.} From a \DU{} perspective, these aspects could lead to users missing key events or dialogue sequences that could affect their ability to follow the plot development. Collecting information at the consumers' end that could help describe and analyse these factors can be done through subjective experiments using surveys and questionnaires~\cite{zhu2015understanding}.

The evolving trends in cinema art (production styles and naturalistic content), technology advances (sound mixing, communication networks, and equipment), and social behaviour (multitasking users and media consumption habits) have impacted the dialogue perception of movies. To keep up with new competitors, involved stakeholders need to consider these factors to deliver high-perceived quality to their consumers and subscribers.

\section{Quantifying \DU{}}\label{sec:tools}
The definition of \DU{} that we propose in this paper is novel. As such, there is no quantitative metric that is designed to measure it. However, each \DU{} factor introduced in the previous section is associated with numerous tools in the literature that can be potentially used. This section provides a comprehensive overview of these tools, linking them to each \DU{} factor (Table~\ref{tab:du-tools}). Pilot experiments were conducted to verify the suitability of these tools over movie-like content. We are particularly interested in seeing which tools are mature enough to be used with real streamed media content, which ones need to be adjusted and tuned to meet streaming context requirements, and which ones are not at all contributing to capturing the \DU{} factors they are targeting.

\subsection{Tools for measuring \DU{}}\label{subsec:tools}

\begin{table*}[tb]
  \centering
  \caption{Mapping tools to \DU{} Factors. Language:l, Accent:a.}
  \resizebox{2\columnwidth}{!}{%
    \begin{tabular}{lr|rrrr|rr|rr|c}
    \hline
    \textbf{Factors} & \multicolumn{1}{c|}{\textbf{Intelligibility Metrics}} & \multicolumn{4}{c|}{\textbf{Quality Metrics}} & \multicolumn{2}{c|}{\textbf{Languages and Accent Classification}} & \multicolumn{2}{c|}{\textbf{Scene Characteristics}} & \textbf{Metadata} \\
          & \multicolumn{1}{c|}{SRMR} & \multicolumn{1}{c}{NISQA-MOS} & \multicolumn{1}{c}{NISQA-noisiness} & \multicolumn{1}{c}{NISQA-intensity} & \multicolumn{1}{c|}{NISQA-colour} & \multicolumn{1}{c}{VoxLingua (L) } & 
          \multicolumn{1}{c|}{CommonAccent (A)} & \multicolumn{1}{c}{P.910-Spatial} & \multicolumn{1}{c|}{P.910-Temporal} &  \\
    \hline
    \textbf{Creation} &    &   &       &       &       &       &       &       &       &  \\
    Artistic/Production Style & \multicolumn{1}{c|}{$\blacksquare$} & \multicolumn{1}{c}{$\blacksquare$} & \multicolumn{1}{c}{$\blacksquare$} & \multicolumn{1}{c}{$\blacksquare$} & \multicolumn{1}{c|}{$\blacksquare$} &   &    & \multicolumn{1}{c}{$\blacksquare$} & \multicolumn{1}{c|}{$\blacksquare$} &  \\
    Language, Accent and Dialogue &       &       &       &       &       & \multicolumn{1}{c}{$\blacksquare$}    & \multicolumn{1}{c|}{$\blacksquare$} &       &       &  \\
    Sound Mix & \multicolumn{1}{c|}{$\blacksquare$} &       &       &       & \multicolumn{1}{c|}{$\blacksquare$} &    &   &       &       & {$\blacksquare$} \\
    \textbf{Delivery} &   &    &       &       &       &       &       &       &       &  \\
    Coding and Streaming & \multicolumn{1}{c|}{$\blacksquare$} & \multicolumn{1}{c}{$\blacksquare$} & \multicolumn{1}{c}{$\blacksquare$} & \multicolumn{1}{c}{$\blacksquare$} & \multicolumn{1}{c|}{$\blacksquare$} &   &    &       &       &  \\
    Equipment and Environment &    &    &       &       &       &       &       &       &       &  \\
    \textbf{Consumption} &    &    &       &       &       &       &       &       &       &  \\
    Social/Immersion/Multitasking &     &    &       &       &       &       &       &       &       &  \\
    \hline
    \end{tabular}%
    }
  \label{tab:du-tools}%
\end{table*}%

\subsubsection{Intelligibility Metrics}\label{subsec:intelli_m}


Computational intelligibility metrics can be divided into intrusive (i.e. use both test and reference signal)~\cite{van2018evaluation} and non-intrusive (i.e. use only the test signal). The former group is not suitable for dialogue movies since a reference signal is missing, and its definition can be ambiguous in this context. Existing non-intrusive intelligibility metrics can be used for the following \DU{} factors: 

\begin{itemize}
    \item \textit{Creation:} artistic/production style, sound mix; 
    \item \textit{Delivery:} coding and streaming.
\end{itemize}


The Speech to Reverberation Modulation energy Ratio (SRMR)~\cite{falk2010non} is a non-intrusive metric designed to assess quality and intelligibility for reverberant speech conditions. The SRMR metric was evaluated in several speech conditions, showing good performance and outperforming intrusive metrics~\cite{falk2015objective}. SRMR can be used to measure the intelligibility of movie dialogues; however, it is advised to use it with caution for certain conditions. For instance, it has been observed that its performance decreases when speech is degraded with multiple impairments such as speech-shaped noise, ideal binary masks, and whenever different signal-to-noise ratios (SNRs) are combined together~\cite{kjems2009role}. This means that the complexity of different sounds that occur in movie scenes, such as foreground music, might mislead SRMR scores. {Collaboration with creators would allow more intrusive testing over dialogue-only tracks, commonly available at a creation stage. This could also be done at a consumer end through object-based audio or ML-based sound separation tools~\cite{shirley2015clean, li2021intelligibility}.}

\subsubsection{Quality Metrics}\label{subsec:quality_m}

Speech quality can also be measured intrusively or non-intrusively. As with intelligibility prediction, using intrusive metrics to predict quality in dialogue movies is not feasible due to the ambiguity that might arise from defining what a reference signal is. Research on non-intrusive speech quality metrics has seen remarkable advances by applying deep learning techniques~\cite{mittag2021nisqa}. These metrics aim at predicting the mean opinion score (MOS), which is the average rating from several human raters for the same media stimulus. MOS predictors can be useful for understanding scenes where dialogue quality varies, which would allow us to investigate the root causes behind quality variations. These causes, as mentioned above, might not necessarily be related to network issues or low-quality codecs but can be due to artistic choices. Quality metrics can be used for the following \DU{} factors:

\begin{itemize}
    \item \textit{Creation:} artistic/production style, sound mix; 
    \item \textit{Delivery:} coding and streaming.
\end{itemize}


The NISQA model~\cite{mittag2021nisqa} is a non-intrusive metric that can be useful to collect features beyond MOS. Besides the MOS prediction, the NISQA model provides measures for noisiness, colouration, discontinuity, and loudness. Being able to separate different quality factors allows us to explore characteristics that go beyond the general MOS value. For instance, speech discontinuity or colouration are traits that can be influenced by artistic choices. Sound mix choices during the creation phase affect dialogue colouration due to timbre interference between speech and other sounds present in the scene. Measuring speech colouration can help to identify where speech complexity could increase due to the sound mix. 

\subsubsection{Scene Characteristics}\label{subsec:scene_c}
How much someone can compress video scenes strongly depends on the content. Content variability can be quantified in terms of spatial and temporal information from video sequences. Such variability will affect the visual perception, which, in turn, will influence the perceived audio quality. Studies have shown that audio quality is strongly influenced by video quality and that video quality tends to dominate the integrated audio-visual QoE~\cite{becerra2021perceptual}. The presence of lip movement has also shown a degree of influence over speech perception. Indeed, it is accepted that speech perception is inherently multimodal~\cite{rosenblum2008speech} with evidence suggesting that sighted humans exploit visual cues such as lip movement when interacting~\cite{rosenblum2008speech}. Scene characteristics can be quantified for the following \DU{} factors:

\begin{itemize}
    \item \textit{Creation:} artistic/production style.
\end{itemize}

ITU-T recommendations P.910 suggest quantifying scene characteristics in terms of spatial and temporal information~\cite{union2008itu}. {The spatial information captures the amount of detail in a single picture, a higher value indicates a more complex scene. Meanwhile, the temporal information measures the temporal changes of a video (a sequence of pictures). A higher value usually corresponds to a high motion video sequence.} Due to their low computational complexity, they are very popular for classifying video databases for encoding or processing purposes.

\subsubsection{Language and Accent Classification}\label{subsec:language_accents}

Open-sourced multilingual automatic speech recognisers (ASR) can help quantify the diversity of languages and accents in movies. For spoken language identification (SLID), the use of deep neural networks has been demonstrated to improve performance and report accurate predictions~\cite{kang2022deep}. In terms of accent identification, several models that deal with this task have been proposed in recent years~\cite{Zheng05}. In multilingual ASR, identifying the accent has been suggested to improve recognition accuracy by mitigating accented speech errors~\cite{Zuluaga23}. In this context, existing language and accent classification methods can be used for the following \DU{} factors:

\begin{itemize}
    \item \textit{Creation:} language, accent and dialogue content.
\end{itemize}

Both language and accent automatic identification have benefited from the rise of new machine learning approaches such as Multi-Task Learning (MTL) and Large Acoustic Models (LAM). LAMs have gained popularity with the implementation of Meta's wav2vec 2.0 Self-Supervised Learning models for speech representation. For language identification, the authors in~\cite{Valk2020VOXLINGUA107AD} presented the VoxLingua model, which was trained using audio segments collected from YouTube videos covering almost 107 languages. In a recent work by Zuluaga-Gomez et al.~\cite{Zuluaga23}, the authors fine-tuned and improved two models: ECAPA-TDNN and wav2vec 2.0 model for accent identification.

\subsubsection{Metadata}\label{subsec:metadata}

The use of metadata to complement signal features has shown promising results for predicting the perceived quality. In~\cite{chinen22_interspeech}, Chinen et al. used the rater group and synthesis system identifiers to complement the feature vector descriptor generated by the wav2vec 2.0 model to predict synthesised speech quality. In terms of \DU{} factors, complementary information regarding the media content could help describe aspects that are not entirely captured with objective methods. For instance, besides information like the movie genre or country of production, which can inform \DU{}, complementary information like the director's name or the movie cast might lead to better descriptors regarding the artistic and production styles that will affect \DU{}. In this study, we recommend using metadata to inform the following \DU{} factors:

\begin{itemize}
    \item \textit{Creation:} artistic/production style, sound mix.
\end{itemize}

{Metadata is transmitted alongside the audio, video, subtitles, and other components of a streamed movie, and its available at the streaming service interface.} The Internet Movie Database (IMDb) is an online dataset containing detailed information for films, TV series, podcasts and streaming content. This information includes the director, production crew, cast, plot summaries, year of release, genre, user ratings/critical reviews, and technical information like sound mix and aspect ratio. This information could help characterise several \DU{} factors like the sound mix format or even use the user's rating to inform about the acceptance of the movie.

\subsection{Quantifying \DU{} features}\label{sec:pilots}

Some of the tools presented in this section were evaluated to verify their suitability for characterising \DU{}. To achieve this, pilot experiments were carried out using available movie clips and non-modified versions of available objective tools. For each experiment, Table~\ref{tab:pilot_exp} contains the \DU{} factor the experiments are associated with, the methods employed, the dataset source and a brief description and technical specifications about the data used in each experiment.

\begin{table*}[hbt!]
  \centering
  \caption{Pilot studies summary. DU: \DU{}.}
  \resizebox{2\columnwidth}{!}{%
    \begin{tabular}{llp{21.07em}lp{35.57em}p{8.93em}}
    \hline
    \textbf{Experiment} & \textbf{DU Factor} & \multicolumn{1}{l}{\textbf{Methods}} & \textbf{Dataset source} & \multicolumn{1}{c}{\textbf{Dataset description}} & \multicolumn{1}{c}{\textbf{Suitability level}} \\
    \hline
    \textbf{Candidate tools} & \multicolumn{1}{p{13.855em}}{Artistic/production style,\newline{}Coding and streaming, \newline{}Sound mix} & Objective metrics\newline{}SRMR (speech intelligibility)\cite{falk2010non}\newline{}NISQA (speech quality)\cite{mittag2021nisqa}\newline{}Spatial/Temporal Index (scene characteristics)\cite{union2008itu}\newline{}Silero VAD (additional feature)\newline{}Google WER (additional feature)\newline{}\newline{}Subjective scores\newline{}Subjects: 3 speech experts\newline{}Material: 4 movie trailer clips\newline{}Equipment: Desktop computer with commercial speakers\newline{}Ranking scale: 0-5 (0 - no dialogue, 1 - bad quiality dialogue, 5 - high quality dialogue) & YouTube & Four YouTube videoclips (movie trailers)\newline{}Audio: stereo, 44.1 kHz sampling rate, 128 kbps bitrate\newline{}Video: MPEG-4, 720p spatial resolution, 23.976 temporal resolution, 4:2:0 color sampling\newline{}Trailers: Casablanca (2 mn 22 s), Spider-Man 3 (2 mn 27 s), The Lord of the Rings-The fellowship of the ring (2 mn 29 s), Slumdog Millionaire (2 mn 15 s) & Mid to low\newline{}-Needs adaptation for movie content \\
    \hline
    \textbf{Language classification} & Language, accent and dialogue & Voxlingua \cite{Valk2020VOXLINGUA107AD}\newline{}107 Languages, 6628 hrs (62 hrs per language) & YouTube & Three YouTube videoclips (10 s segmented movie trailers)\newline{}Audio: stereo, 44.1 kHz sampling rate, 128 kbps bitrate\newline{}Video: MPEG-4, 720p spatial resolution, 23.976 temporal resolution, 4:2:0 color sampling\newline{}Trailers: Narcos (11 clips, English, Spanish), Squid Game (12 clips, Korean), 1899 (11 clips, English, Spanish, French, German) & Low\newline{}-Needs adaptation for movie content\newline{}-Add source separation + VAD component \\
    \textbf{Accent classification} & Language, accent and dialogue & CommonAccent \cite{Zuluaga23}\newline{}Accents: 16 English (African, Australia, Bermuda, Canada, England, Hong Kong, Indian, Ireland, Malaysia, New Zealand, Philippines, Scotland, Singapore, South Atlantic, US, and Wales), 4 German, 6 Spanish, and 5 Italian & YouTube & Two YouTube videoclips (segmented movie trailers)\newline{}Audio: stereo, 44.1 kHz sampling rate, 126 kbps bitrate\newline{}Video: Advance video codec, 640x360 spatial resolution, 24 temporal resolution, 4:2:0 color sampling\newline{}Trailers: Sense8 (7 clips, 2 to 9 s), John Wick 4 (6 clips, 3 to 14 s) & Low\newline{}-Needs adaptation for movie content\newline{}-Add source separation + VAD component \\
    \hline
    \end{tabular}%
  \label{tab:addlabel}%
    }
  \label{tab:pilot_exp}%
\end{table*}%

\subsubsection{Measuring \DU{}: Candidate Tools}\label{subsec:pilot_tools}
In this experiment, we test some instrumental objective metrics (presented in Section~\ref{subsec:tools}) for \DU{}. We try to explore how these metrics, which were developed to capture specific aspects from speech/audio and video, will perform over movie-type content. {The films were selected to represent different film styles, genres, languages and contemporaneity (year of release). Although we try to represent different types of content, the main target is to gather information on how the available tools perform over movie-like content.} Overall, ten different features were extracted to measure aspects affecting \DU{}. {The audio WAV files were segmented into clips of 10~s duration. Objective measurements were computed for each segment separately and they were aligned later for each segment. Three (3) expert listeners (with normal hearing) were presented with the four (4) movie trailers (only once) and they were asked to rate the \DU{} levels. The listening environment was kept the same for each listening experimental session.} Details regarding the movie clips, the metrics tested, and the collection of subjective data are presented in Table~\ref{tab:pilot_exp}.

\begin{figure}
    \centering
    \includegraphics[width=0.48\textwidth]{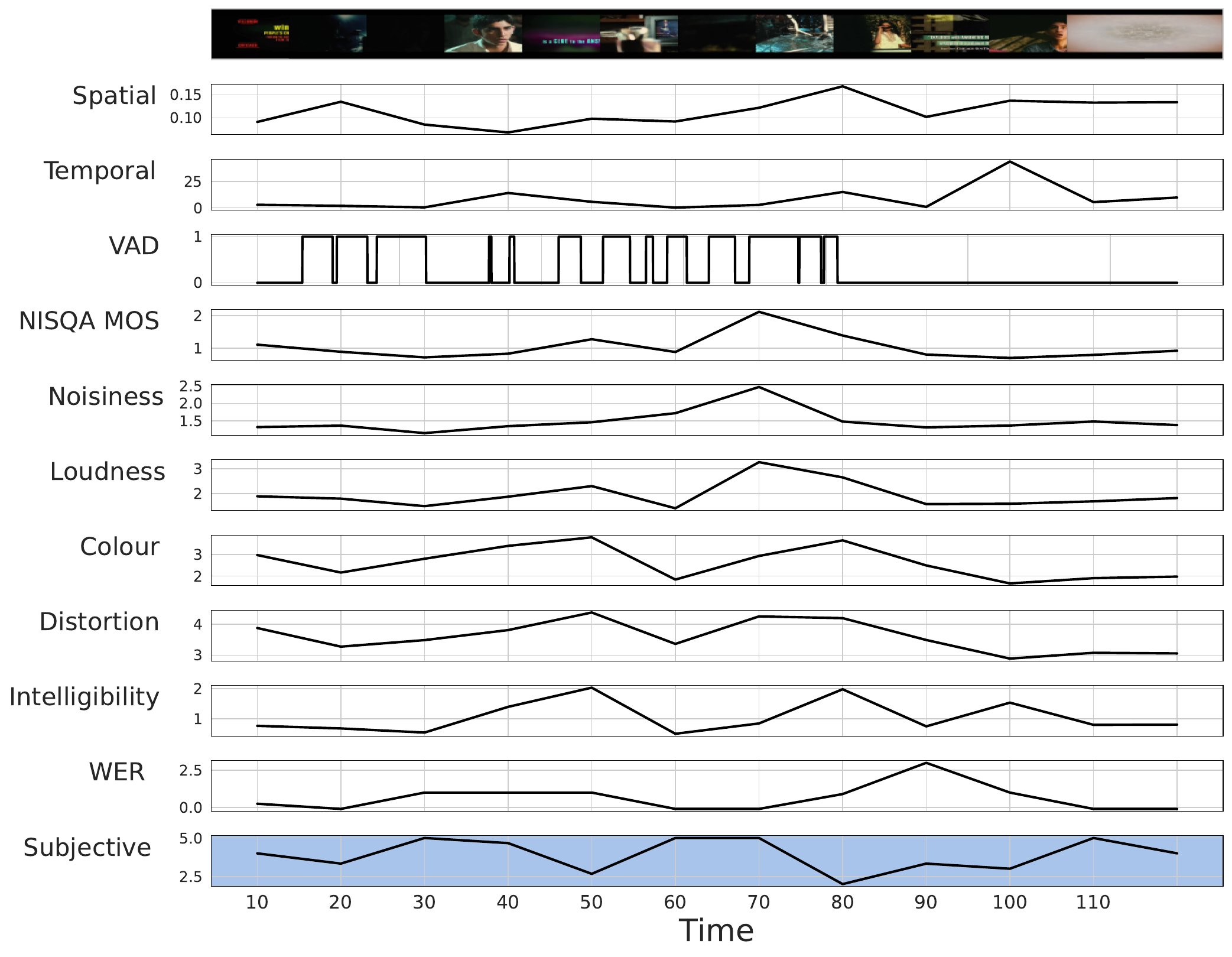}\\
    \caption{Feature analysis for movie trailers. Slumdog Millionaire (Danny Boyle, Loveleen Tandan, 2008).} 
    \label{fig:tools_DU}
\end{figure}

Figure~\ref{fig:tools_DU} presents the report for ten objective measurements plus the subjective scores for one movie clip. This representative report aims to show how these features vary in relation to a subjective \DU{} score. Results from this experiment suggest that, by themselves, these objective measurements are not able to fully characterise the perceived \DU{}. However, variations in their scores, which were also observed among the subjective ratings, indicate that they are capturing variations in certain characteristics related to the perceived \DU{}. Further experiments, including more human participants and a curated dataset, are necessary to confirm that, in fact, these features have the potential to be used to feed a predictive model for \DU{}.

\subsubsection{Language identification}
\label{subsec:language_id}

This experiment seeks to evaluate the application of language identification models over movie-like content. {Language identification could potentially help the implementation of personalised assistance tools. For instance, speech enhancement and smart captioning whenever a specific language is detected.} We are interested in evaluating the suitability of available tools for the purpose of \DU{} characterisation. In this experiment, we evaluate the pre-trained version of the VoxLingua model~\cite{Valk2020VOXLINGUA107AD} over audio samples extracted from YouTube video clips. {Three video clips were selected to represent different languages over streamed TV content (e.g., English, German, Spanish, Korean). The audio component of the clips was extracted and segmented into 10 s sequences.} Details for the clip samples can be found in Table~\ref{tab:pilot_exp}.

Results from this experiment showed the difficulty of the VoxLingua to correctly predict some languages, such as English, Welsh, and Dutch. Interestingly, the model predicted Latin and Javanese languages when there was no speech activity in the given audio segment (background music and sound effects). This highlights the need to add a VAD component (plus source separation for overlapping sound) to avoid incongruous outcomes. Although the model showed good performance over controlled speech segments~\cite{Valk2020VOXLINGUA107AD}, the results from our experiment suggest that the model is sensitive to more challenging segments that include noisy and competing sound effects.

\subsubsection{Accent identification}

For this experiment, we tested the CommonAccent pre-trained model presented in~\cite{Zuluaga23}. The model implementation takes an audio sample and returns percentage values for the detected accents. The model was tested over YouTube clips selected for their diversity of English accents in their dialogues. {The audio component for the two clips was extracted and segmented into clips of 2 to 14 s duration.} Details for the movie clips and their content can be found in Table~\ref{tab:pilot_exp}. Results showed the model's difficulty in identifying the target accent in movies. Moreover, the acoustic and linguistic information (e.g. accent portrayal of the actor, and movie script) also has a high effect on the accuracy of the accent identification. Although the model was able to recognise UK and US English accents, the model tended to force prediction when no dialogue was present or when the target accent was outside the 16 accents from which the system was trained.

\subsubsection{Modelling \DU{}}
\begin{figure}
    \centering
    \includegraphics[width=0.48\textwidth]{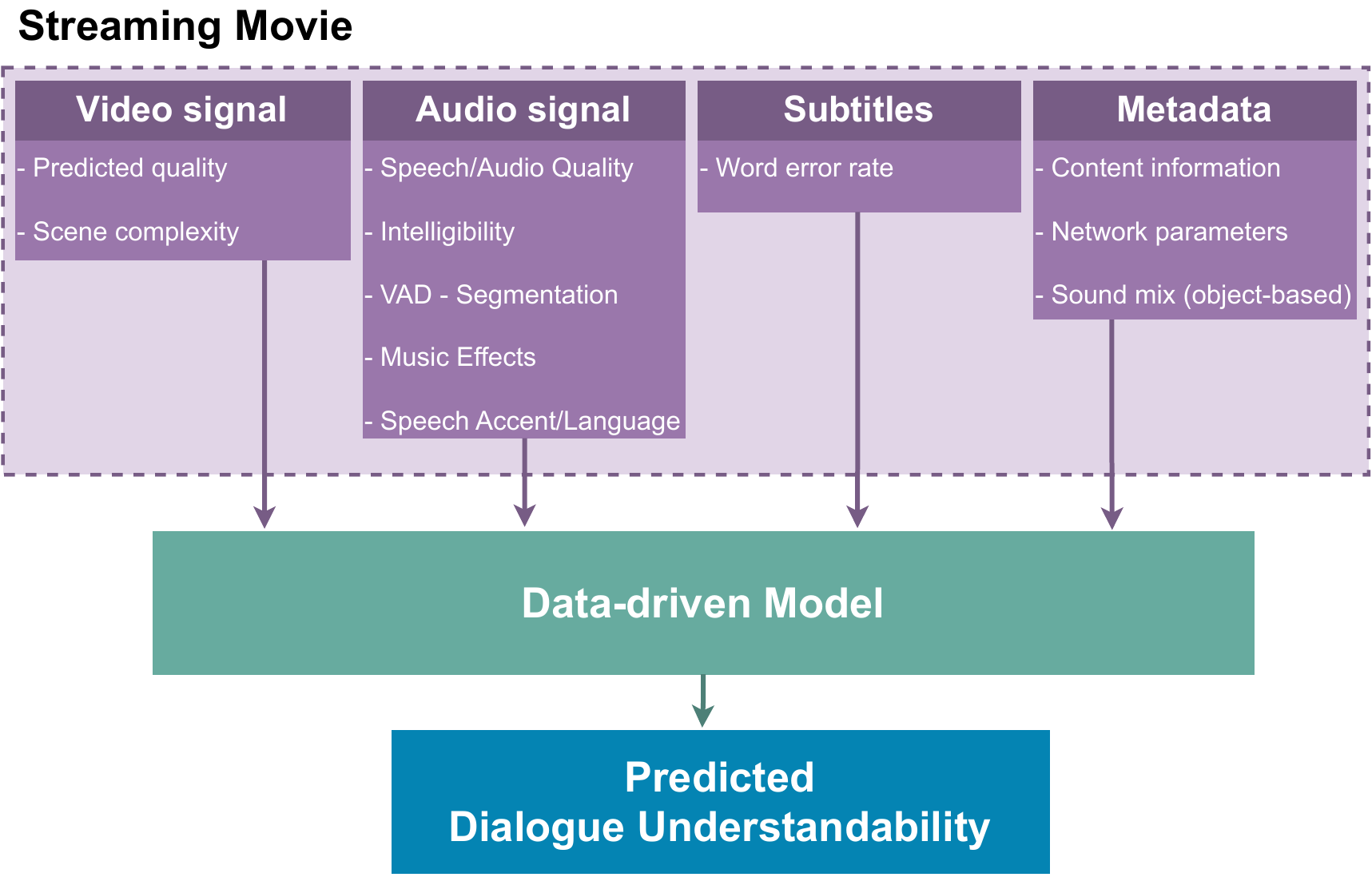}\\
    \caption{Measuring \DU{}. Feature sources and potential metrics to predict \DU{}.} 
    \label{fig:modelling_DU}
\end{figure}

{Results from pilot experiments showed that current methods had trouble performing over movie-like audio clips. These results confirm the need for adaptation and possibly additional tools (e.g., VAD component plus source separation) before using them over movie-like content. Having looked at the suitability of the different tools, Figure \ref{fig:modelling_DU} presents a working diagram for modelling \DU{}. The diagram shows the main components of a streaming movie (e.g., video, audio, subtitles, metadata) that could act as potential sources to extract features describing \DU{}. The extracted features would act as input for a data-driven model trained to predict the levels of \DU{} for streaming movies. It is important to note that, although an approximation of the perceived \DU{} could be attained using this type of model, it still won't be able to capture the entire range of factors affecting it. The diagram only presents what is currently feasible, considering the available material and tools. A large portion of information, mainly from the creation and consumption stages, is still needed to fully measure the perceived \DU{}. In that sense, this diagram should be taken as an initial approximation of what modelling \DU{} constitutes.}

\section{Discussion}\label{sec:discussion}

\subsection{Sound mix format evolution}


{The timeline presented in Figure~\ref{fig:evolution_mix} shows how sound mix formats have evolved and produced new formats that target a more realistic and immersive audio experience. Multichannel-based formats have dominated the cinema sound landscape and they were the standard for several cinema and home theatre settings until the 2010s. The introduction of the Auro-3D format as an initial three-dimensional audio experience attempt (height sound channels), served as base to the development for object-based formats which allow audio elements to move freely around the listener in a three-dimensional space. More recently, AI-based formats are being developed trying to adjust their setup to match how the listener would perceive the sound~\cite{nam2021ai}. This timeline shows the efforts of the sound cinema industry to create more immersive formats, yet; the downmixing process, where these formats are adapted to fit different devices, has received less attention. The design of these devices (e.g., thinner TVs and smaller speakers), their functional characteristics (e.g., smartphones plus headphones), and the diversity of devices (e.g., smartphones, personal computers, smart TVs, gaming consoles) could affect the resulting sound mix perceived by users consuming movies through such devices. Overall, we can observe an increase in sound mixing formats, especially in the last 25 years. Although some of them have not been widely adopted, such diversity still poses a challenge to sound mixing engineers, streaming service companies and hardware manufacturers wherever they need to decide to implement sound mixing adaptations for different devices and consumption settings.}

\subsection{Utility of \DU{}}

In this paper, we presented a working definition for \DU{} and a list of factors influencing it. A \DU{} perspective, which relates to the QoE framework and the movie lifecycle stages, facilitates the analysis and understanding of extent and effectiveness of the methods used to improve \DU{} such as dialogue enhancement, sound mix, and smart subtitles.

Dialogue enhancement is an active area of research that aims to enhance the dialogue sounds in movie content~\cite{master2023deepspace}. As with other speech-related fields, it has seen great advances by applying more sophisticated deep-learning-based approaches (e.g., MTLs, LAMs)~\cite{article67}. {This method can be implemented in the delivery stage by streaming service providers targeting the consumption stage where users could customise their listening preferences.} More recently, Amazon Prime's \textit{Dialogue Boost}~\cite{amazonboost2023} was released as an accessibility tool allowing users to change the relative speech volume with respect to the background music and effects. Subjective experiments, designed and informed by \DU{}, are needed to understand how Amazon Prime's \textit{Diaologue Boost}, and other dialogue enhancement methods, are affecting \DU{}. For instance, reducing the background noise, music, etc., in a scene undoubtedly affects the context or tone of a scene at the consumption stage. With the advent of dialogue boosting, to what level artistic choices in a scene be influenced in the creation stage?

For the sound mix process, object-based audio proposes an alternative to the traditional channel-based way of mixing sound and gives an opportunity to improve broadcast accessibility for hard-of-hearing individuals~\cite{shirley2015clean}. This approach proposes to transmit audio signals as speech components and mix them at the receiver's end-point using descriptive metadata. This allows more flexibility whenever a component needs to be prioritised based on the user's demand. From a \DU{} perspective, this approach would be beneficial for managing sound elements that are considered critical for the dialogue and plot understanding. In addition, it would allow a more personalised solution based on individual needs instead of a unique enhancement process that tries to cover a wide range of users and consumption settings (e.g., devices, environments, user traits, etc.). {The adoption of object-based audio formats has been growing in recent years, especially among streaming service providers (e.g., Netflix, Amazon Prime, Disney+, AppleTV). This is thanks to the implementation of the delivery systems required to handle object-based audio transmission. There are, however, some aspects that need to be explored whenever object-based audio is used, for instance, an audio-based mixing approach would remove the artistic control from creators and rely on other stakeholders (e.g., hardware manufacturers, streaming service providers, or end consumers) to get the ‘right’ sound mix. How would this affect the choices made by the creators?}

Subtitles help individuals interpret and grasp content that might otherwise be inaccessible to them. {Production companies and streaming service providers delegate the generation of subtitles to in-house production teams, external subtitling agencies and expert freelancers. Despite being considered a key tool for dialogue and content understanding, potential negative effects such as scene blocking, font size and text legibility, and demanding levels of visual attention are reported by users~\cite{crabb2015online}. {Moreover, subtitles may not be equally effective for everyone. Reading difficulties arising from different levels of quality of education and socio-economic or neurological factors (e.g., dyslexia) could impact the use of subtitles, increasing the cognitive load and reducing content engagement for consumers.} One alternative is the implementation of smart subtitling (or dynamic captioning). This approach aims to optimise the position of captions through speaker recognition and scene analysis~\cite{brown2015dynamic}. Early studies have shown good levels of acceptance among users reporting more immersive experiences and less missed content during the viewing sessions~\cite{brown2015dynamic}. There is, however, the need to explore to what level the subtitling positioning would impact artistic choices like the scene composition and the cinematography design. \DU{} provides the needed framework to explore and understand such impact. Subjective experiments, guided by \DU{}, will help understand and guide the design of smart subtitles with minimal impact on the intended content.}

\subsection{Characterising \DU{}}


This study provides a formal categorization of the different aspects that influence how we consume movies outside cinema theatres. We have presented a list of potential methods to characterise \DU{}, and they were tested over a set of pilot studies to explore their applicability and maturity to measure or describe \DU{} factors. We analyse the potential of these tools to characterise \DU{} based on their availability, practicality, and maturity.

From an availability standpoint, tools to automatically measure factors like equipment and environment, social immersion, and multitasking were not readily available for testing. Although there are studies focused on measuring these factors, they are mostly restricted to subjective experiments~\cite{hines2015visqolaudio}. In terms of practicality, mapping the \DU{} factors in the streaming pipeline and linking them to the involved stakeholders allowed us to understand the degree of control they offer. Creation factors like the artistic and production style, the language accent and dialogue content and the sound mix cannot be avoided. However, through a \DU{} approach we are in the capacity of using the available information (metadata) and automatic tools to detect potential issues and enforce methods to mitigate their effect over \DU{} (e.g., dialogue enhancement, subtitles). Regarding their maturity, the pilot studies allowed us to explore the suitability of these methods for movie-type content. Based on those results, it was concluded that the available tools can't be used to measure \DU{} factors in their current state. Complementary tools and model adaptation (e.g., ASR components, fine-tuning steps) are needed to capture movie-like content characteristics. In addition, larger experiments are necessary to evaluate these adapted tools.

\subsection{Limitations and Future Experiments}

Merely identifying the origin of a factor to a particular stage or stakeholder will not rectify the \DU{}. Systematically understanding the factors contributing to \DU{} facilitates an engaged conversation about the extent to which each stakeholder can contribute to solutions. The potential and limitations of technology-based solutions are also framed, acknowledging that there are no universal silver bullet solutions. The list and the mapping presented in this study is an attempt to cover the entire spectrum of factors that influence \DU{} at different stages, but it does not answer the question of what is the share of the impact that these factors have on the overall \DU{}. The set of pilot experiments presented in this study served to evaluate the level of maturity of currently available measurement tools. Future experiments, to quantify the impact of these factors, will require larger content material curated following a \DU{} perspective. A promising initiative is the dataset presented at~\cite{hung2022large}, containing 1600 h of audio tracks from movie and TV shows for speech and music activity detection. The dataset is manually annotated with time stamps for speech and music segments in the audio track. Datasets like this could be used for the implementation and evaluation of \DU{} optimization tools.

\section{Conclusions}\label{sec:conclusion}

In this study, we provided a working definition of \DU{}. The term describes the evolving factors that shape the way we consume movies and streamed-like content. We organised these factors into six sub-categories based on their root causes. We linked them to existing QoE influential factors, the common streaming lifecycle stages, and the involved stakeholders to bring attention to how manageable these factors are and who are the main responsible actors that they depend on. The results from a set of pilot studies, that verified the suitability of the current methods in the literature, highlighted the need for adaptation before using them over movie-like content. They also evidenced that certain factors, being artistic and aesthetic aspects decided by content creators, cannot be avoided. But, they can be spotted at an early stage, and help guide the implementation of assistance tools to mitigate their effect and improve \DU{} (e.g., dialogue enhancement, smart subtitles, or sound mix). Our study can help the design of relevant datasets, experimental protocols, and objective tools to characterise the perceived \DU{} of movie-like content.

We begin this paper by asking the question, \textit{why are we streaming movies with subtitles?} The answer lies in the evolving complexity of processes for movie creation, transmission and consumption. More and more subtitles are being used in response to a poor \DU{} experience. Our work provides a systematic structure for researchers to explore and implement tools to improve every viewer's experience.

\bibliographystyle{ieeetr}
\bibliography{du}

\end{document}